\documentclass[11pt,a4paper]{article}
\pdfoutput=1
\usepackage{jheppub}
\usepackage{url}
\usepackage{tikz}
\usepackage{subcaption}
\DeclareMathOperator{\Tr}{Tr}

\newcommand{\cob}{\delta}

\newcommand{\vep}{\varepsilon}

\newcommand{\hf}{\frac{1}{2}}
\newcommand{\qu}{\frac{1}{4}}
\newcommand{\til}[1]{\widetilde{#1}}

\newcommand{\del}{\partial}

\newcommand{\bra}{\langle}
\newcommand{\ket}{\rangle}

\newcommand{\ka}{\kappa}
\newcommand{\h}[1]{\widehat{#1}}
\newcommand{\bt}{\beta}
\newcommand{\ga}{\gamma}

\newcommand{\rt}[1]{\sqrt{#1}}
\newcommand{\cO}{\mathcal{O}}
\newcommand{\cZ}{\mathcal{Z}}
\newcommand{\cF}{\mathcal{F}}
\newcommand{\cR}{\mathcal{R}}

\newcommand{\cH}{\mathcal{H}}

\newcommand{\cW}{\mathcal{W}}

\begin{document}

\title{Quenched free energy from spacetime D-branes}

\author{Kazumi Okuyama}

\affiliation{Department of Physics, Shinshu University,\\
3-1-1 Asahi, Matsumoto 390-8621, Japan}

\emailAdd{kazumi@azusa.shinshu-u.ac.jp}

\abstract{
We propose a useful integral representation of the quenched free energy
which is applicable to any random systems.
Our formula involves the generating function of multi-boundary correlators,
which can be interpreted on the bulk gravity side 
as spacetime D-branes
introduced by Marolf and Maxfield in [arXiv:2002.08950].
As an example, we apply our formalism to the Airy limit of the
random matrix model and compute its quenched free energy
under certain approximations of the generating function
of correlators.
It turns out that the resulting quenched free energy
is a monotonically decreasing function of the temperature, as expected.
}

\maketitle


\section{Introduction}\label{sec:intro}
Recent works suggest that the gravitational path integral generically
includes the
contribution of Euclidean wormholes 
and the result should be interpreted as a certain
ensemble average,
at least for lower dimensional gravities in spacetime dimensions $d\leq3$
(see e.g. \cite{Saad:2018bqo,Saad:2019lba,Marolf:2020xie,Cotler:2020ugk}).
The most well-known example is the holographic duality
between Sachdev-Ye-Kitaev (SYK) model \cite{Sachdev,kitaev2015simple}
and Jackiw-Teitelboim (JT) gravity in two dimensions 
\cite{Jackiw:1984je,Teitelboim:1983ux}.
In a remarkable paper \cite{Saad:2019lba}, it was shown that 
the path integral of JT gravity
is equivalent to a doubled scaled random matrix model
and the genus expansion of this matrix model describes 
the splitting/joining of baby universes.

In a system with random Hamiltonian, one can define
two kinds of free energy:
the quenched free energy
$\bra\log Z\ket$ and the annealed free energy $\log\bra Z\ket$,
where $Z$ is the partition function
with the inverse temperature $\bt=1/T$
\begin{equation}
\begin{aligned}
 Z\equiv Z(\bt)=\Tr e^{-\bt H}, 
\end{aligned} 
\end{equation}
and the bracket $\bra\cdot\ket$ denotes the ensemble average.
In a random system, thermodynamic quantities like the free energy $F$ 
and the entropy $S$ should be defined via the quenched free energy \footnote{In this paper, we will loosely use the name ``quenched free energy''
for both $\bra\log Z\ket$ and $F=-T\bra\log Z\ket$ depending on the context.
We hope this will not cause a confusion to the readers.}
\begin{equation}
\begin{aligned}
 F=-T\bra\log Z\ket,\quad S=-\frac{\del F}{\del T},
\end{aligned} 
\end{equation}
which are expected to satisfy various thermodynamic inequalities.
In particular, the positivity of entropy $S\geq0$ requires
that the quenched free energy is a monotonically decreasing function of temperature,
while the annealed free energy $F_{\text{ann}}=-T\log\bra Z\ket$ is
not necessarily a monotonic function of temperature.

In a recent paper \cite{Engelhardt:2020qpv},
the quenched free energy in JT gravity is studied by the replica method
\begin{equation}
\begin{aligned}
 \bra\log Z\ket=\lim_{n\to0}\frac{\bra Z^n\ket-1}{n},
\end{aligned} 
\label{eq:replica}
\end{equation}
and it is argued that the wormholes connecting different replicas, so-called
replica wormholes, play an important role 
(see also \cite{Johnson:2020mwi,Alishahiha:2020jko}).\footnote{In a different 
context of the computation of von Neumann entropy
of Hawking radiation, replica wormholes also play
an essential role to recover the unitary Page curve
\cite{Penington:2019kki,Almheiri:2019qdq}.
See also \cite{Almheiri:2020cfm} for a nice review on this subject.}
However, the replica computation in \cite{Engelhardt:2020qpv}
does not give rise to a well-behaved monotonic free energy at low temperature
due to the limitation of the approximation used in \cite{Engelhardt:2020qpv}.
More importantly, as emphasized in \cite{Engelhardt:2020qpv}
there is a fundamental problem in the replica computation of the quenched free
energy since the analytic continuation of $\bra Z^n\ket$ from a positive integer
$n$ to $n=0$ is highly non-unique.

The replica method \eqref{eq:replica} is just a convenient trick to
compute the quenched free energy and one can in principle
compute it by directly taking the average of the quantity $\log Z$.
This is demonstrated in our previous paper \cite{Okuyama:2020mhl}
in a simple example of the Gaussian matrix model.
We computed the average $\bra\log Z\ket$ directly by the matrix integral
at finite $N$ and found that the resulting quenched free energy is 
a monotonic function of the temperature, as expected.
It should be emphasized that if we take the average $\bra\log Z\ket$ directly
without using the replica method \eqref{eq:replica}, the result is uniquely
determined and there is no ambiguity in the computation. However,
the direct computation in \cite{Okuyama:2020mhl} relies
on the property of the finite $N$ matrix model and 
it is not straightforward to generalize it to 
the double scaled matrix model of JT gravity.
Thus, it is desirable to find a general technique to compute the quenched free energy
without resorting to the replica trick \eqref{eq:replica}.

In this paper, we propose a useful integral representation of the quenched free energy
\begin{equation}
\begin{aligned}
 \bra \log Z\ket=\log\bra Z\ket-\int_0^\infty\frac{dx}{x}\Bigl[\big\bra e^{-Zx}\big\ket-e^{-\bra Z\ket x}\Bigr],
\end{aligned} 
\label{eq:formula}
\end{equation}
which can be applied to any random system.
As far as we know, this expression of quenched free energy \eqref{eq:formula}
has not appeared in the literature before.\footnote{A similar, 
but slightly different expression of
$\bra\log Z\ket$ is discussed in \cite{svaiter2016distributional}.}
Our formula \eqref{eq:formula} follows from the simple identity \eqref{eq:ab-int}
and the derivation of \eqref{eq:formula} does not rely on any approximation.
In section \ref{sec:replica}, we will argue that 
our formula \eqref{eq:formula} can be obtained from
the replica method \eqref{eq:replica}
under a certain prescription of the
analytic continuation of $\bra Z^n\ket$.

Our formula \eqref{eq:formula}
might have an interesting bulk gravity interpretation.
As discussed in \cite{Marolf:2020xie},
the insertion of $Z$ into the expectation value $\bra\cdot\ket$
corresponds to adding an extra boundary to the spacetime
in the bulk gravity picture.
The insertion of $e^{-Zx}$ creates many boundaries in spacetime
and it is identified in \cite{Marolf:2020xie} as the ``spacetime D-brane'' (or SD-brane for short). Based on this picture, our formula \eqref{eq:formula}
suggests that the bulk gravity dual of the quenched free energy
involves a superposition of SD-branes.
Here, by ``superposition'' we mean that we have to integrate over the parameter
$x$ in the SD-brane operator $e^{-Zx}$.

Introducing the generating function $\cZ(x)$ of the 
connected correlators $\bra Z^n\ket_c$
\begin{equation}
\begin{aligned}
 \cZ(x)=-\sum_{n=1}^\infty\frac{(-x)^n}{n!}\bra Z^n\ket_c,
\end{aligned} 
\label{eq:cZ-def}
\end{equation}
\eqref{eq:formula} is written as
\begin{equation}
\begin{aligned}
 \bra \log Z\ket=\log\bra Z\ket-\int_0^\infty\frac{dx}{x}\Bigl[
e^{-\cZ(x)}-e^{-\bra Z\ket x}\Bigr],
\end{aligned} 
\label{eq:formula2}
\end{equation}
where we used the relation $\bra e^{-Zx}\ket=e^{-\cZ(x)}$.
Thus the problem boils down to the computation of the
generating function $\cZ(x)$ in \eqref{eq:cZ-def}.
As a simple application of our formalism \eqref{eq:formula2},
we consider the quenched free energy in the Airy limit 
of Gaussian matrix model.\footnote{See e.g. \cite{Ginsparg:1993is} 
for a review of the
Airy limit of matrix model.}
We find that at genus-zero the
generating function $\cZ(x)$ can be written in a closed form in terms of the
Lambert function.
If we plug this genus-zero approximation
of $\cZ(x)$ into \eqref{eq:formula2}, we find a monotonically decreasing behavior
of the quenched free energy
as a function of the temperature.
However, if we include higher genus corrections to
$\cZ(x)$ in \eqref{eq:formula2} it leads to an unphysical diverging behavior
of free energy as $T\to0$.
This simply means that the genus expansion is not a good approximation
at low temperature, and hence we will consider a different 
approximation of $\cZ(x)$ in \eqref{eq:cZ-lowsum} which is  
applicable at low temperature.
Then, we find a monotonic behavior of the quenched free energy
in this low temperature approximation \eqref{eq:cZ-lowsum}. 

This paper is organized as follows.
In section \ref{sec:formula}, we explain the derivation of the formula
\eqref{eq:formula} for the quenched free energy. We also see that 
this expression \eqref{eq:formula} can be obtained from the replica
method \eqref{eq:replica}. Then we briefly comment on the possible
bulk gravity interpretation of our formula \eqref{eq:formula}.
In section \ref{sec:Airy}, we apply our formalism \eqref{eq:formula2}
to the Airy limit of Gaussian matrix model.
We find that the genus expansion of $\cZ(x)$ in the Airy limit is written in terms
of the Lambert function. The appearance of the Lambert function
can be understood from the string equation
in a shifted background
due to the insertion of the operator $e^{-Zx}$.
We find a monotonic behavior of the quenched free energy in a certain
approximation of $\cZ(x)$.
Finally, we conclude in section \ref{sec:conclusion}
with some discussion for the future problems.
In appendix \ref{app:high},
we compute the high temperature expansion of the quenched free energy in the Airy limit.

\section{General formula for the quenched free energy}\label{sec:formula}

\subsection{Direct derivation of \eqref{eq:formula}}
Let us prove our general formula \eqref{eq:formula} for the quenched free energy.
\eqref{eq:formula} is based on the following simple relation
\begin{equation}
\begin{aligned}
 -\int_0^\infty\frac{dx}{x}(e^{-ax}-e^{-bx})=\log\frac{a}{b}.
\end{aligned} 
\label{eq:ab-int}
\end{equation}
Note that, if we take the first term  of \eqref{eq:ab-int}
only, then the integral is
logarithmically divergent 
\begin{equation}
\begin{aligned}
 -\int_{\vep}^\infty\frac{dx}{x}e^{-ax}=\log(a\vep)+\ga+\cO(\vep),
\end{aligned} 
\end{equation}
where $\vep$ is a small regularization parameter and $\ga$ is the Euler-Mascheroni constant.
However, if we compute the difference of two such integrals as in \eqref{eq:ab-int}
the divergence from $x=0$ is canceled and we get the finite result \eqref{eq:ab-int}.
Setting $a=Z$ and $b=\bra Z\ket$ in \eqref{eq:ab-int} we obtain
\begin{equation}
\begin{aligned}
 -\int_0^\infty\frac{dx}{x}\Big(e^{-Zx}-e^{-\bra Z\ket x}\Big)=\log\frac{Z}{\bra Z\ket}.
\end{aligned} 
\label{eq:Zlog-int}
\end{equation}
Finally, taking the expectation value of the both sides of
\eqref{eq:Zlog-int} and using the relation $\bra 1\ket=1$, we find
our desired result \eqref{eq:formula}.
We emphasize that our formula \eqref{eq:formula}
is an exact relation and we did not use any approximation 
in deriving \eqref{eq:formula}.

Let us consider an alternative derivation of \eqref{eq:formula}
by expanding the quenched free energy around the annealed free energy.
\begin{equation}
\begin{aligned}
 \bra\log Z\ket&=\big\bra\log\big(\bra Z\ket+Z-\bra Z\ket\big)\big\ket\\
&=\log\bra Z\ket+\Biggl\bra\log\Biggl(1+\frac{Z-\bra Z\ket}{\bra Z\ket}\Biggr)
\Biggr\ket\\
&=\log\bra Z\ket-\sum_{k=1}^\infty\frac{(-1)^{k}}{k}\frac{\big\bra\big(Z-\bra Z\ket\big)^k\big\ket}{\bra Z\ket^k}.
\end{aligned} 
\label{eq:sumk}
\end{equation}
Then using the relation
\begin{equation}
\begin{aligned}
 \int_0^\infty \frac{dy}{y} e^{-y}y^{k}=(k-1)!,\qquad(k\geq1),
\end{aligned} 
\label{eq:yk-int}
\end{equation}
we rewrite the summation in \eqref{eq:sumk} as 
\begin{equation}
\begin{aligned}
 -\sum_{k=1}^\infty\frac{(-1)^{k}}{k}\frac{\big\bra\big(Z-\bra Z\ket\big)^k\big\ket}{\bra Z\ket^k}&=
-\int_0^\infty \frac{dy}{y}e^{-y}\sum_{k=1}^\infty\frac{(-y)^{k}}{k!}\frac{\big\bra\big(Z-\bra Z\ket\big)^k\big\ket}{\bra Z\ket^k}\\
&=-\int_0^\infty \frac{dy}{y}e^{-y}\left[\Biggl\bra
\exp\Biggl(-y\frac{Z-\bra Z\ket}{\bra Z\ket}\Biggr)\Biggr\ket-1\right]\\
&=-\int_0^\infty \frac{dy}{y}
\left[\Biggl\bra
\exp\Biggl(-y\frac{Z}{\bra Z\ket}\Biggr)\Biggr\ket-e^{-y}\right]
\end{aligned} 
\end{equation}
Finally, by rescaling the integration variable $y=\bra Z\ket x$ we arrive at
our formula \eqref{eq:formula}.

\subsection{Relation to the replica method}\label{sec:replica}
Let us consider a derivation of our formula \eqref{eq:formula}
from the replica method \eqref{eq:replica}.
In the replica method, we need to analytically continue
the $n$-point function $\bra Z^n\ket$ from a positive integer $n$ to $n=0$.
As pointed out in \cite{Engelhardt:2020qpv},
the analytic continuation of $\bra Z^n\ket$ is not unique
and there might be some ambiguity
in the computation of the right hand side of \eqref{eq:replica}.
It turns out that there is a natural definition of
the analytic continuation of $\bra Z^n\ket$ which leads to our formula
\eqref{eq:formula}.

In general, the correlator $\bra Z^n\ket$ is expanded as a combination of 
the connected correlators $\bra Z^k\ket_{c}$.
If we order this expansion by the number of connected components,
the first term is the totally disconnected part $\bra Z\ket^n$.
The next term is $\bra Z\ket^{n-2}\bra Z^2\ket_{c}$
and the coefficient of this term is $\binom{n}{2}=\hf n(n-1)$
which is the number of ways to choose two boundaries out of $n$ boundaries.
Then $\bra Z^n\ket$ is expanded as
\begin{equation}
\begin{aligned}
 \bra Z^n\ket&=\bra Z\ket^n+\hf n(n-1)\bra Z\ket^{n-2}\bra Z^2\ket_{c}+\cdots\\
&=\bra Z\ket^n\left[1+\hf n(n-1)\frac{\bra Z^2\ket_c}{\bra Z\ket^2}+\cdots\right].
\end{aligned} 
\label{eq:binom}
\end{equation}
Taking the $n\to0$ limit in \eqref{eq:replica}, we obtain
\begin{equation}
\begin{aligned}
 \bra \log Z\ket=\log\bra Z\ket-\hf \frac{\bra Z^2\ket_c}{\bra Z\ket^2}+\cdots.
\end{aligned} 
\label{eq:example-z2}
\end{equation}
In this computation, up to the overall factor
$\bra Z\ket^n$ the $n$-dependence in \eqref{eq:binom}
is essentially a polynomial in $n$
and  hence the analytic continuation can be defined unambiguously.

To see the general structure, it is convenient to define
\begin{equation}
\begin{aligned}
 w_n=\frac{\bra Z^n\ket_{c}}{\bra Z\ket^n}.
\end{aligned} 
\label{eq:wn}
\end{equation}
Then the expansion \eqref{eq:binom}
is generalized as (see e.g. \cite{Okuyama:2019xvg})
\begin{equation}
\begin{aligned}
 \bra Z^n\ket=\bra Z\ket^n\sum_{j_i\geq0~(i=2,3,\cdots)}\frac{n!}{(n-\sum_{l\geq2} lj_l)!}
\prod_{k\geq2} \frac{1}{j_k!}\left(\frac{w_k}{k!}\right)^{j_k}.
\end{aligned} 
\label{eq:young}
\end{equation}
Now using the relation
\begin{equation}
\begin{aligned}
 \lim_{n\to0}\frac{1}{n}\frac{n!}{(n-m)!}=(-1)^{m-1}(m-1)!,
\end{aligned} 
\label{eq:binom-cont}
\end{equation}
the quenched free energy becomes
\begin{equation}
\begin{aligned}
 \bra \log Z\ket=\log \bra Z\ket+\sum_{\substack{j_i\geq0~(i=2,3\cdots)\\
(j_2,j_3,\cdots)\ne(0,0,\cdots)}}
(-1)^{\sum_{l\geq2} lj_l-1}\Bigl(\sum_{l\geq2} lj_l-1\Bigr)!
\prod_{k\geq2}\frac{1}{j_k!} \left(\frac{w_k}{k!}\right)^{j_k}.
\end{aligned} 
\end{equation}
Again using the trick \eqref{eq:yk-int} this is rewritten as
\begin{equation}
\begin{aligned}
 \bra \log Z\ket&=\log \bra Z\ket-\sum_{\substack{j_i\geq0~(i=2,3,\cdots)\\
(j_2,j_3,\cdots)\ne(0,0,\cdots)}}
\int_0^\infty \frac{dy}{y} e^{-y} (-y)^{\sum_{l\geq2} lj_l}
\prod_{k\geq2}\frac{1}{j_k!} \left(\frac{w_k}{k!}\right)^{j_k}\\
&=\log \bra Z\ket-\int_0^\infty \frac{dy}{y}e^{-y}\left[\exp\Biggl(
\sum_{k=2}^\infty (-y)^k\frac{w_k}{k!}\Biggr)-1\right]\\
&=\log \bra Z\ket-\int_0^\infty \frac{dx}{x}
\Biggl[\exp\Biggl(\sum_{k=1}^\infty \frac{(-x)^k}{k!}\bra Z^k\ket_c\Biggr)
-e^{-\bra Z\ket x}\Biggr],
\end{aligned}
\label{eq:replica-que}
\end{equation}
where in the last step we changed the integration variable $y=\bra Z\ket x$.
One can see that the first term of the integrand in \eqref{eq:replica-que} 
is $e^{-\cZ(x)}$ defined in \eqref{eq:cZ-def}, and hence 
\eqref{eq:replica-que} agrees with our formula
\eqref{eq:formula2}.

To summarize, we have derived our formula \eqref{eq:formula}
from the replica method by using the prescription
\eqref{eq:binom-cont} for the analytic continuation.
For instance, the example in \eqref{eq:example-z2} of this computation 
corresponds to the $m=2$ case of \eqref{eq:binom-cont}
\begin{equation}
\begin{aligned}
 \lim_{n\to0}\frac{1}{n}\frac{n!}{(n-2)!}=\lim_{n\to0}\frac{1}{n} n(n-1)=-1.
\end{aligned} 
\end{equation} 
At fixed $m$, the factor $\frac{n!}{(n-m)!}$ in \eqref{eq:binom-cont}
is a polynomial in $n$
\begin{equation}
\begin{aligned}
 \frac{n!}{(n-m)!}=n(n-1)\cdots (n-m+1),
\end{aligned} 
\end{equation}
and hence we believe that our prescription \eqref{eq:binom-cont}
of the analytic continuation is a natural choice.

\subsection{Possible interpretation of \eqref{eq:formula}}
Our formula \eqref{eq:formula} 
is very suggestive for a possible interpretation of the bulk spacetime
picture of the quenched free energy.
In \cite{Marolf:2020xie}, it is discussed that the operator
$e^{-Zx}$ corresponds to the so-called ``spacetime D-brane'' (SD-brane).
Our formula \eqref{eq:formula} suggests that the bulk spacetime picture of
quenched free energy involves a superposition of SD-branes.
As mentioned in section \ref{sec:intro},
in order to obtain the quenched free energy,
we have to integrate over the parameter $x$ in the SD-brane operator
$e^{-Zx}$.

Recently, in \cite{Penington:2019kki,Almheiri:2019qdq} 
the von Neumann entropy of Hawking radiation is
computed by the replica method and it is argued that the so-called replica wormholes 
are essential for recovering the unitary Page curve.
In the saddle point approximation of gravitational path integral, the dominant contribution comes from either
maximally connected or maximally disconnected spacetimes
and these contributions exchange dominance around the Page time.
On the other hand, in our formula \eqref{eq:formula} 
no such saddle point approximation is made.
In our derivation of \eqref{eq:formula} from the replica method,
all terms in the decomposition of $\bra Z^n\ket$
in \eqref{eq:young} contribute to
the quenched free energy and they add up to the generating function
$\bra e^{-Zx}\ket$. The first term $\log\bra Z\ket$ in \eqref{eq:formula}
comes from the maximally disconnected part $\bra Z\ket^n$, while
the remaining term in \eqref{eq:formula} is a contribution of SD-brane
$e^{-Zx}$
which does not have a simple interpretation as the contribution
of maximally connected part.
As we will see in section \ref{sec:constitutive},
the insertion of the operator $e^{-Zx}$ can be interpreted as a shift of the background couplings, or
a deformation of the matrix model potential.

It is argued in \cite{Engelhardt:2020qpv} that
the contribution of the so-called replica wormholes should be included in
the replica computation of the quenched free energy.
The second term of \eqref{eq:formula} represents the deviation from
the annealed free energy, which might be interpreted as the  
contribution of replica wormholes. It is interesting that
even after taking the limit $n\to0$ \eqref{eq:replica},
infinite number of boundaries contributes to $\bra e^{-Zx}\ket=e^{-\cZ(x)}$ in
\eqref{eq:cZ-def}. In other words, 
we need to include a coherent superposition $\cZ(x)$ of 
the infinite number of boundaries for the computation of quenched free energy. 

In general,
the large $N$ limit serves as a semi-classical approximation in the random matrix model.
One might think that we can see an exchange of dominance between disconnected and
connected contributions in the large $N$ limit.
However, if we naively take the large $N$ limit at fixed temperature,
the disconnected correlator  always dominates
over the connected 
correlators. Around the temperature $T\sim N^{-2/3}$, the disconnected
and connected contributions become comparable \cite{Okuyama:2019xvg}, but this temperature is pushed to zero as $N\to\infty$ which implies that the disconnected part 
is always dominant at finite temperature.
As discussed in \cite{Engelhardt:2020qpv,Johnson:2020mwi},
to focus on this temperature scale $T\sim N^{-2/3}$ 
one can take a certain double scaling limit, called 
the Airy limit of random matrix model, which we will consider in the next section.

\section{Quenched free energy in the Airy limit}\label{sec:Airy}
In this section we apply our general formula \eqref{eq:formula}
to the Airy limit of Gaussian random matrix model.
In this case, the connected part of the $n$-point correlator
$\bra Z^n\ket_c$ is known in the integral representation \cite{okounkov2002generating}.
For instance, $\bra Z^n\ket_c$ for $n=1,2,3$ can be written in the closed form
\cite{okounkov2002generating,Beccaria:2020ykg}
\begin{equation}
\begin{aligned}
 \bra Z(\bt)\ket&=\frac{e^{\frac{\xi}{12}}}{\rt{4\pi\xi}},\\
\bra Z(\bt)^2\ket_c&=\bra Z(2\bt)\ket\text{Erf}\Biggl(\rt{\frac{\xi}{2}}\Biggr),\\
\bra Z(\bt)^3\ket_c&=\bra Z(3\bt)\ket\Biggl[1-12T\Bigl(\rt{3\xi},\frac{1}{\rt{3}}\Bigr)\Biggr],
\end{aligned}
\label{eq:airy-conn} 
\end{equation}
where $\text{Erf}(z)$ is the error function and $T(z,a)$
denotes the Owen's $T$-function defined by
\begin{equation}
\begin{aligned}
 T(z,a)=\frac{1}{2\pi}\int_0^a dt \frac{e^{-\frac{z^2}{2}(1+t^2)}}{1+t^2}.
\end{aligned} 
\end{equation}
The parameter $\xi$ in \eqref{eq:airy-conn} is given by
\begin{equation}
\begin{aligned}
 \xi=\hbar^2\bt^3,
\end{aligned} 
\label{eq:xi}
\end{equation}
where $\hbar$ is the genus-counting parameter.
In the Airy limit, $\hbar$ and $\bt$ appear in $\bra Z^n\ket_c$
only through the combination $\xi$.

In principle one can compute $\bra Z^n\ket_c$
using the integral representation of \cite{okounkov2002generating}, but
it is not straightforward to
find the generating function $\cZ(x)$ of the connected correlators
since the integral in \cite{okounkov2002generating} 
is difficult to compute in a closed form for general $n$.
In fact, the closed form expression of $\bra Z^n\ket_c$ is not known for $n\geq4$.
Instead, here we will consider the genus expansion
of the generating function $\cZ(x)$
\begin{equation}
\begin{aligned}
 \cZ(x)=\sum_{g=0}^\infty \xi^{g-1}\cZ_g(x),
\end{aligned} 
\label{eq:cZ-genus}
\end{equation}
where $\cZ_g(x)$ is the generating function of
the genus-$g$ part of the connected correlators.
It turns out that $\cZ_g(x)$ is written as a combination of the Lambert $W$-function.
If we use the genus-zero approximation
$\cZ(x)\approx \xi^{-1}\cZ_0(x)$ in \eqref{eq:formula2},
we find that the quenched free energy
$F=-T\bra\log Z\ket$ decreases monotonically as a function of temperature.
However, the genus expansion is not a good approximation
at low temperature since the expansion parameter
$\xi$ in \eqref{eq:cZ-genus} becomes large when $T\ll \hbar^{2/3}$.

At low temperature we can use the result of \cite{Okuyama:2019xvg}
that the connected part of the $n$-point correlator
is approximated by the one-point function\footnote{Strictly speaking, this
result \eqref{eq:Znc-approx} is shown in the 
Gaussian matrix model \cite{Okuyama:2019xvg}, 
but we expect that \eqref{eq:Znc-approx} holds in the Airy limit as well.
}
\begin{equation}
\begin{aligned}
 \bra Z(\bt)^n\ket_c\approx\bra Z(n\bt)\ket,\quad(T\lesssim \hbar^{2/3}).
\end{aligned} 
\label{eq:Znc-approx}
\end{equation}
One can see this behavior explicitly from the exact result
of $\bra Z(\bt)^n\ket_c$ for $n=2,3$ in \eqref{eq:airy-conn}. We emphasize that
\eqref{eq:Znc-approx} is the general property of the connected correlator for any $n$.
Then $\cZ(x)$ at low temperature is approximated as
\begin{equation}
\begin{aligned}
 \cZ(x)\approx-\sum_{n=1}^\infty\frac{(-x)^n}{n!}\bra Z(n\bt)\ket.
\end{aligned} 
\label{eq:cZ-lowsum}
\end{equation}
The right hand side of this equation can be computed explicitly
since the exact form of the one-point function $\bra Z(n\bt)\ket$
is known in the Airy limit. 
Plugging \eqref{eq:cZ-lowsum} into \eqref{eq:formula2}, we find
the monotonic behavior of the quenched free energy even at low temperature
$T\lesssim \hbar^{2/3}$.

\subsection{Quenched free energy at genus-zero}
Let us consider the genus-zero part of the expansion of
$\cZ(x)$ in \eqref{eq:cZ-genus}.
The connected correlator
$\bra Z^n\ket_c$ is obtained by acing the ``boundary creation operator''
$\h{Z}$ to the free energy
$\cF(\{t_k\})$ of 2d topological gravity \cite{Moore:1991ir}
\begin{equation}
\begin{aligned}
 \bra Z^n\ket_c=\big(\h{Z}\big)^n \cF,
\end{aligned} 
\label{eq:corr-top}
\end{equation}
where $\h{Z}$ is given by
\begin{equation}
\begin{aligned}
 \h{Z}=g_s\rt{\frac{\bt}{2\pi}}\sum_{k=0}^\infty \bt^k\del_k
\end{aligned} 
\label{eq:hatZ}
\end{equation}
with $\del_k\equiv\frac{\del}{\del t_k}$. 
Then the generating function $\cZ(x)$ is written as
\begin{equation}
\begin{aligned}
 \cZ(x)=-\sum_{n=1}^\infty \frac{(-x)^n}{n!}\big(\h{Z}\big)^n \cF
=\cF-e^{-\h{Z}x}\cF.
\end{aligned} 
\label{eq:cZ-shift1}
\end{equation}
$g_s$ in \eqref{eq:hatZ} is 
the natural genus-counting parameter in the topological gravity 
\begin{equation}
\begin{aligned}
 \cF=\sum_{g=0}^\infty g_s^{2g-2}\cF_g,
\end{aligned}
\label{eq:cF-genus} 
\end{equation}
and $g_s$ is related to $\hbar$ in \eqref{eq:xi} by
\begin{equation}
\begin{aligned}
 g_s=\rt{2}\hbar.
\end{aligned} 
\end{equation}

The Airy limit corresponds to
the trivial background $t_n=0~(n\geq0)$.
As demonstrated in \cite{Zograf:2008wbe,Okuyama:2019xbv,Okuyama:2020ncd},
in order to compute the connected correlators \eqref{eq:corr-top}
one can utilize the KdV equation obeyed by the specific heat $u=g_s^2\del_0^2\cF$
\cite{Witten:1990hr,Kontsevich:1992ti}.
In the KdV approach of the computation of $\bra Z^n\ket_c$, it is convenient to turn on $t_0,t_1$
\begin{equation}
\begin{aligned}
 t_0,t_1\ne0,\quad t_n=0~~(n\geq2),
\end{aligned} 
\label{eq:off-shell}
\end{equation}
and set $t_0=t_1=0$ at the end of the calculation.
One can show that in this subspace \eqref{eq:off-shell} 
$u$ is given by
\begin{equation}
\begin{aligned}
 u=\frac{t_0}{1-t_1},
\end{aligned} 
\label{eq:u01}
\end{equation}
without any higher genus corrections.

At genus-zero, the general expression of the connected correlator \eqref{eq:corr-top}
reduces to \cite{Moore:1991ir}
\begin{equation}
\begin{aligned}
 \bra Z^n\ket_c^{g=0}=\left(\frac{\bt}{2\pi}\right)^{\frac{n}{2}}
(g_s\del_0)^{n-2}\frac{e^{n\bt u_0}}{n\bt},
\end{aligned} 
\label{eq:Zn-g0}
\end{equation}
where $u_0$ is determined by the genus-zero string equation \cite{Itzykson:1992ya}
\footnote{The all-genus string equation is given by
\begin{equation}
\begin{aligned}
 u=\sum_{k=0}^\infty t_k\cR_k,
\end{aligned} 
\end{equation}
where $\cR_k=\frac{u^k}{k!}+\cO(\del_0)$ is the Gelfand-Dikii differential polynomial
determined by the recursion relation
\begin{equation}
\begin{aligned}
 (2k+1)\del_0\cR_{k+1}=\frac{\hbar^2}{2}\del_0^3\cR_k+2u\del_0\cR_k+(\del_0u)\cR_k.
\end{aligned} 
\end{equation}
}
\begin{equation}
\begin{aligned}
 u_0=\sum_{k=0}^\infty t_k\frac{u_0^k}{k!}.
\end{aligned} 
\label{eq:string}
\end{equation}
In our case \eqref{eq:off-shell}, $u_0$ is equal to $u$ in \eqref{eq:u01} 
\begin{equation}
\begin{aligned}
 u_0=\del_0^2\cF_0=\frac{t_0}{1-t_1}.
\end{aligned} 
\label{eq:u0-airy}
\end{equation}
Plugging this $u_0$ into \eqref{eq:Zn-g0}
the genus-zero part of the $n$-point correlator
$\bra Z^n\ket_c^{g=0}$ becomes
\begin{equation}
\begin{aligned}
 \bra Z^n\ket_c^{g=0}=\frac{(1-t_1)^{2-n}}{2\xi}n^{n-3}\left(\frac{\xi}{\pi}\right)^{\frac{n}{2}}e^{\frac{n\bt t_0}{1-t_1}},
\end{aligned} 
\label{eq:Zg0-t0t1}
\end{equation}
where $\xi$ is defined in \eqref{eq:xi}. Then, by setting $t_0=t_1=0$ we find
$\bra Z^n\ket_c^{g=0}$ in the Airy limit 
\begin{equation}
\begin{aligned}
 \bra Z^n\ket_c^{g=0}=\frac{1}{2\xi}n^{n-3}\left(\frac{\xi}{\pi}\right)^{\frac{n}{2}}.
\end{aligned} 
\label{eq:Zng0-Airy}
\end{equation}

Now let us consider the generating function of the genus-zero correlators
in \eqref{eq:Zng0-Airy}
\begin{equation}
\begin{aligned}
 \xi^{-1}\cZ_0(x)=-\sum_{n=1}^\infty \frac{(-x)^n}{n!}
\bra Z^n\ket_c^{g=0}=\frac{1}{2\xi}\sum_{n=1}^\infty 
\frac{(-1)^{n-1}n^{n-3}}{n!}
\left(x\rt{\frac{\xi}{\pi}}\right)^n.
\end{aligned} 
\label{eq:Z0-sum}
\end{equation}
It turns out that this summation is closely related to the
Lambert function $W(z)$
\begin{equation}
\begin{aligned}
 W(z)=\sum_{n=1}^\infty \frac{(-1)^{n-1}n^{n-1}}{n!}z^n,
\end{aligned} 
\label{eq:w-sum}
\end{equation}
which satisfies the relation
\begin{equation}
\begin{aligned}
 z=W(z)e^{W(z)}.
\end{aligned} 
\end{equation}
From \eqref{eq:Z0-sum} and \eqref{eq:w-sum},
one can see that $\cZ_0$ obeys
\begin{equation}
\begin{aligned}
 (z\del_z)^2\cZ_0=\hf W(z),
\end{aligned} 
\label{eq:dif-Z0}
\end{equation}
where $z$ is related to $x$ by
\begin{equation}
\begin{aligned}
 z=x\rt{\frac{\xi}{\pi}}.
\end{aligned} 
\label{eq:z-onshell}
\end{equation}
Using the property of the Lambert function
\begin{equation}
\begin{aligned}
 z\del_zW(z)=\frac{W(z)}{1+W(z)},
\end{aligned} 
\label{eq:W-dif}
\end{equation}
\eqref{eq:dif-Z0} is integrated as
\begin{equation}
\begin{aligned}
 \cZ_0=\hf\Biggl(\frac{1}{6}W^3+\frac{3}{4}W^2+ W\Biggr).
\end{aligned} 
\label{eq:cZ0}
\end{equation}
Here we have suppressed the argument of $W(z)$ for brevity.

Then at genus-zero the quenched free energy \eqref{eq:formula2}
is written as 
\begin{equation}
\begin{aligned}
 \bra \log Z\ket^{g=0}=\log \bra Z\ket^{g=0}
-\int_0^\infty\frac{dx}{x}\Bigl[e^{-\xi^{-1}\cZ_0(x)}-e^{-\bra Z\ket^{g=0}x}\Bigr],
\end{aligned} 
\label{eq:que-g0}
\end{equation}
where $\bra Z\ket^{g=0}=\frac{1}{\rt{4\pi\xi}}$ is the genus-zero one-point function.
Using the fact that $z\in[0,\infty]$ is mapped to $W\in[0,\infty]$,
we can take $W$ as the integration variable in \eqref{eq:que-g0}.
From \eqref{eq:W-dif}, \eqref{eq:que-g0} is rewritten as
\begin{equation}
\begin{aligned}
  \bra \log Z\ket^{g=0}=-\hf\log(4\pi\xi)
-\int_0^\infty\frac{dW}{W}(1+W)\Bigl[e^{-\frac{1}{2\xi}\left(
\frac{1}{6}W^3+\frac{3}{4}W^2+W\right)}-e^{-\frac{1}{2\xi}We^W}\Bigr].
\end{aligned} 
\label{eq:que-g0-Wint}
\end{equation}

\begin{figure}[htb]
\centering
\includegraphics[width=0.45\linewidth]{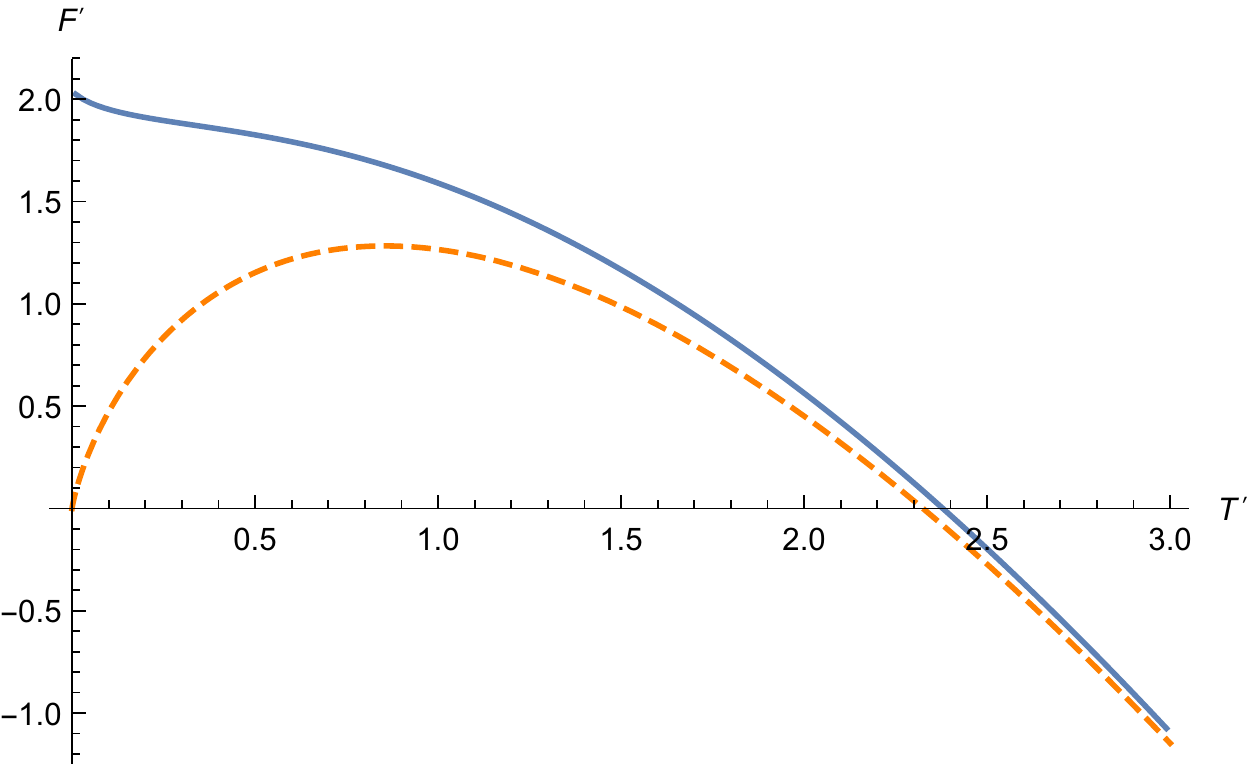}
  \caption{
Plot 
of the genus-zero free energy  for the Airy case.
The vertical and the horizontal axes are the rescaled free energy $F'=\hbar^{-2/3}F$
and the rescaled temperature
$T'=\hbar^{-2/3}T$, respectively.
The blue solid curve represents the quenched free energy 
in \eqref{eq:que-g0-Wint}
while the orange dashed curve is the annealed free energy in \eqref{eq:Fann-g0}.
}
  \label{fig:F0-Airy}
\end{figure}

In Fig.~\ref{fig:F0-Airy} we show the plot of rescaled free energy 
$F'=\hbar^{-2/3}F$ of the Airy case
at genus-zero. In other words, $F'$ is defined by
\begin{equation}
\begin{aligned}
 F'=-T'\bra \log Z\ket,\quad T'=\hbar^{-2/3}T,\quad
\xi=T'^{-3}.
\end{aligned} 
\label{eq:F'T'}
\end{equation}
From  Fig.~\ref{fig:F0-Airy} one can  
see that the quenched free energy at genus-zero 
(blue solid curve) is a monotonic function
of the temperature.
As a comparison, we have also plotted the annealed free energy
at genus-zero (orange dashed curve in Fig.~\ref{fig:F0-Airy})
\begin{equation}
\begin{aligned}
 F_{\text{ann}}'=-T'\log\bra Z\ket^{g=0}=\hf T'\log(4\pi T'^{-3}),
\end{aligned}
\label{eq:Fann-g0} 
\end{equation}
which is not a monotonic function of the temperature.

\subsection{Higher genus corrections}
Next let us consider the higher genus corrections $\cZ_g$ in
\eqref{eq:cZ-genus}.
As we will see below, $\cZ_g$ can be systematically obtained from
the KdV equation. Our starting point is the
KdV equation for $u$
\begin{equation}
\begin{aligned}
 \del_1u=u\del_0u+\frac{\hbar^2}{6}\del_0^3u.
\end{aligned} 
\label{eq:KdV-u}
\end{equation}
Introducing the potential $\phi$ by
\begin{equation}
\begin{aligned}
 u=\del_0\phi,
\end{aligned} 
\end{equation}
\eqref{eq:KdV-u} is integrated once as
\begin{equation}
\begin{aligned}
 \del_1\phi=\hf(\del_0\phi)^2+\frac{\hbar^2}{6}\del_0^3\phi.
\end{aligned} 
\label{eq:phi-eq}
\end{equation}
Note that $\phi$ is written as
\begin{equation}
\begin{aligned}
 \phi=g_s^2\del_0\cF=2\hbar^2\del_0\cF.
\end{aligned} 
\label{eq:phi-cF}
\end{equation}

We can derive 
an equation obeyed by $\cZ(x)$ by acting the 
operator $e^{-\h{Z}x}$ on both sides of \eqref{eq:phi-eq}.
From \eqref{eq:cZ-shift1} and \eqref{eq:phi-cF}, $e^{-\h{Z}x}\phi$ is written as
\begin{equation}
\begin{aligned}
 e^{-\h{Z}x}\phi=2\hbar^2\del_0 e^{-\h{Z}x}\cF=2\hbar^2\del_0(\cF-\cZ)=\phi-2\hbar\cW,
\end{aligned} 
\end{equation}
where we defined
\begin{equation}
\begin{aligned}
 \cW=\hbar\del_0\cZ.
\end{aligned} 
\label{eq:def-cW}
\end{equation}
Acting $e^{-\h{Z}x}$ on both sides of \eqref{eq:phi-eq}, we find
\begin{equation}
\begin{aligned}
 \del_1(\phi-2\hbar\cW)=\hf(\del_0\phi-2\hbar\del_0\cW)^2+\frac{\hbar^2}{6}
\del_0^3(\phi-2\hbar\cW).
\end{aligned} 
\label{eq:phi-eZ}
\end{equation}
From \eqref{eq:phi-eq} and \eqref{eq:phi-eZ}, we arrive at the equation
for $\cW$ \footnote{This equation was originally found by Kazuhiro Sakai.
The author would like to thank him for sharing his unpublished note.}
\begin{equation}
\begin{aligned}
 \del_1\cW=u\del_0\cW-\hbar(\del_0\cW)^2+\frac{\hbar^2}{6}\del_0^3\cW.
\end{aligned} 
\label{eq:eq-cW}
\end{equation}

Let us consider the genus expansion of $\cW$
\begin{equation}
\begin{aligned}
 \cW=\bt^{-\hf}\sum_{g=0}^\infty\xi^{g-\hf}\cW_g.
\end{aligned} 
\end{equation}
It turns out that $\cW_g$ is a rational function of $W\equiv W(z)$ with $z$
given by \eqref{eq:z-onshell}. To compute $\cW_g$ recursively from \eqref{eq:eq-cW},
it is useful to work on the subspace \eqref{eq:off-shell}.
On this space \eqref{eq:off-shell},
one can easily show that $z$ in \eqref{eq:z-onshell} is generalized as
(see \eqref{eq:Zg0-t0t1})
\begin{equation}
\begin{aligned}
 z=\frac{1}{1-t_1}x e^{\frac{\bt t_0}{1-t_1}}\rt{\frac{\xi}{\pi}}.
\end{aligned} 
\label{eq:z-offshell}
\end{equation}
Then the derivative $\del_{0,1}$ appearing in \eqref{eq:eq-cW}
is written in terms of $D=z\del_z$.
We also note that $\hbar$ and $t_1$ appear only via the combination
\begin{equation}
\begin{aligned}
 \hat{\hbar}\equiv\frac{\hbar}{1-t_1}.
\end{aligned} 
\label{eq:hat-hbar}
\end{equation}
With these remarks in mind, $\del_0\cW$ and $\del_1\cW$ at $t_0=t_1=0$ 
are given by
\begin{equation}
\begin{aligned}
 \del_0\cW\Big|_{t_0=t_1=0}&=\bt D\cW,\\
\del_1\cW\Big|_{t_0=t_1=0}&=(D+\hbar\del_{\hbar})\cW.
\end{aligned} 
\end{equation}
Then after setting $t_0=t_1=0$ \eqref{eq:eq-cW} reduces to
\begin{equation}
\begin{aligned}
 (D+2g-1)\cW_g=\frac{1}{6}D^3\cW_{g-1}-\sum_{h=0}^{g}D\cW_hD\cW_{g-h}.
\end{aligned} 
\label{eq:recg}
\end{equation}
From the definition of $\cW$ \eqref{eq:def-cW}, 
one can easily show that $\cZ_g$ and $\cW_g$ are related by
\begin{equation}
\begin{aligned}
 \cW_g=D\cZ_g.
\end{aligned} 
\end{equation}
To solve \eqref{eq:recg} recursively, we rewrite \eqref{eq:recg} as
\begin{equation}
\begin{aligned}
 \Bigl[(1+2D\cW_0)D+2g-1\Bigr]\cW_g=\frac{1}{6}D^3\cW_{g-1}-\sum_{h=1}^{g-1}
D\cW_hD\cW_{g-h}.
\end{aligned} 
\end{equation}
Note that, from \eqref{eq:W-dif} $D=z\del_z$  is written as
\begin{equation}
\begin{aligned}
 D=\frac{W}{1+W}\del_W,
\end{aligned} 
\label{eq:def-D}
\end{equation}
and from \eqref{eq:dif-Z0} $D\cW_0$ is given by
\begin{equation}
\begin{aligned}
 D\cW_0=D^2\cZ_0=\hf W.
\end{aligned} 
\label{eq:DW0}
\end{equation}
Finally we arrive at the recursion relation for $\cW_g$
\begin{equation}
\begin{aligned}
 (W\del_W+2g-1)\cW_g=\frac{1}{6}D^3\cW_{g-1}-\sum_{h=1}^{g-1}D\cW_hD\cW_{g-h}.
\end{aligned} 
\label{eq:Wg-rec}
\end{equation}
One can easily solve this relation recursively starting from $D\cW_0$ in \eqref{eq:DW0}.
More explicitly, \eqref{eq:Wg-rec} is solved as
\begin{equation}
\begin{aligned}
 \cZ_g=\int dW (1+W)W^{-2g}\int dW W^{2g-2}
\left[\frac{1}{6}D^3\cW_{g-1}-\sum_{h=1}^{g-1}D\cW_hD\cW_{g-h}\right].
\end{aligned} 
\end{equation}
Using this formalism, we can compute $\cZ_g$ up to any desired order.
For instance,
the first few terms are given by
\begin{equation}
\begin{aligned}
 \cZ_1&
=\frac{1}{24}\log(1+W),\\
\cZ_2&=\frac{5W-19W^2+4W^3}{2880(1+W)^5},\\
\cZ_3&=\frac{35 W-1977 W^2+12393
   W^3-15919 W^4+3960 W^5-16
   W^6}{725760 (1+W)^{10}}.
\end{aligned} 
\label{eq:Zg-result}
\end{equation} 

Using this result \eqref{eq:Zg-result},
we can study the behavior of free energy including the higher genus corrections up to
genus-$g$
\begin{equation}
\begin{aligned}
 \bra\log Z\ket_{[g]}=\log\bra Z\ket_{[g]}-\int_0^\infty\frac{dW}{W}(1+W)
\left[e^{-\cZ_{[g]}}-e^{-\bra Z\ket_{[g]}\rt{\frac{\pi}{\xi}}We^W}\right],
\end{aligned} 
\label{eq:que-Z[g]}
\end{equation}
where we defined $\bra Z\ket_{[g]}$ and $\cZ_{[g]}$ by
\begin{equation}
\begin{aligned}
 \bra Z\ket_{[g]}&=\frac{1}{\rt{4\pi\xi}}\sum_{n=0}^{g} \frac{\xi^n}{12^n n!},\\
\cZ_{[g]}&=\sum_{h=0}^g\xi^{h-1}\cZ_h.
\end{aligned} 
\label{eq:Z[g]}
\end{equation}

\begin{figure}[htb]
\centering
\subcaptionbox{Free energy up to $g=1$ \label{sfig:F1}}{\includegraphics[width=0.3\linewidth]{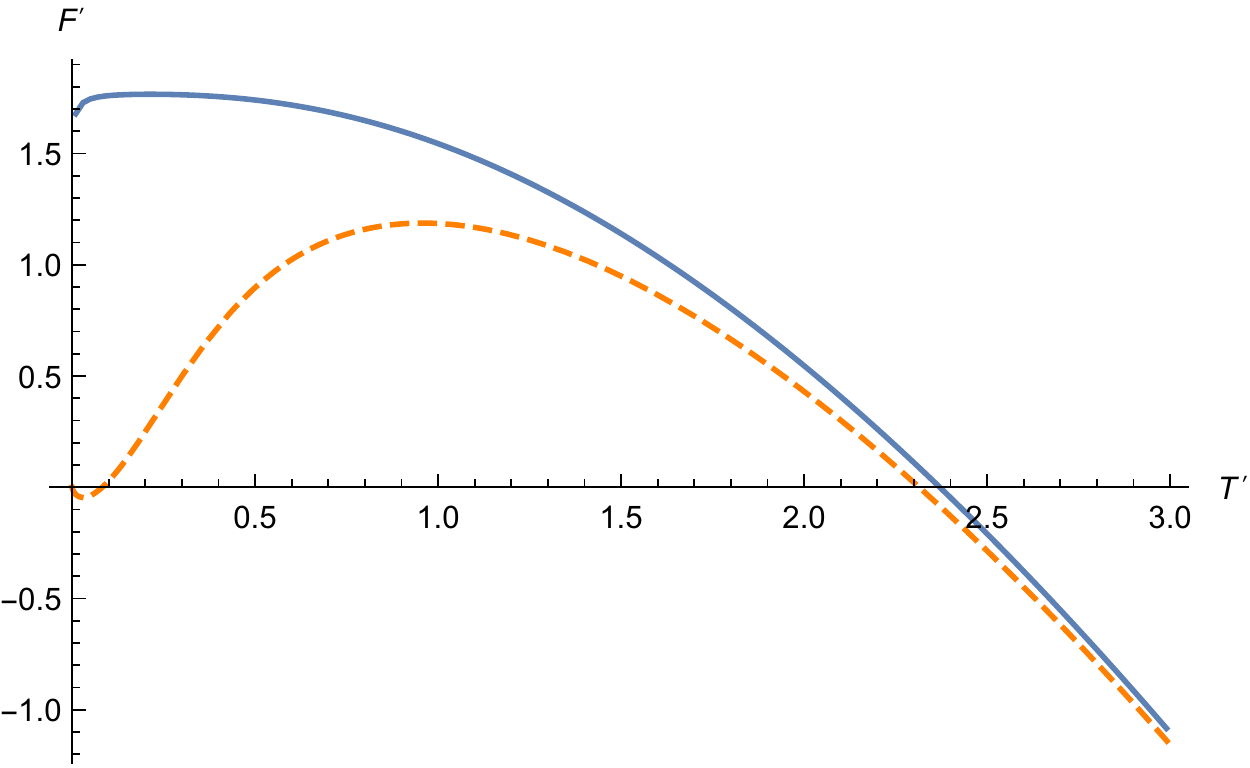}}
\hskip5mm
\subcaptionbox{Free energy up to $g=2$ \label{sfig:F2}}{\includegraphics[width=0.3\linewidth]{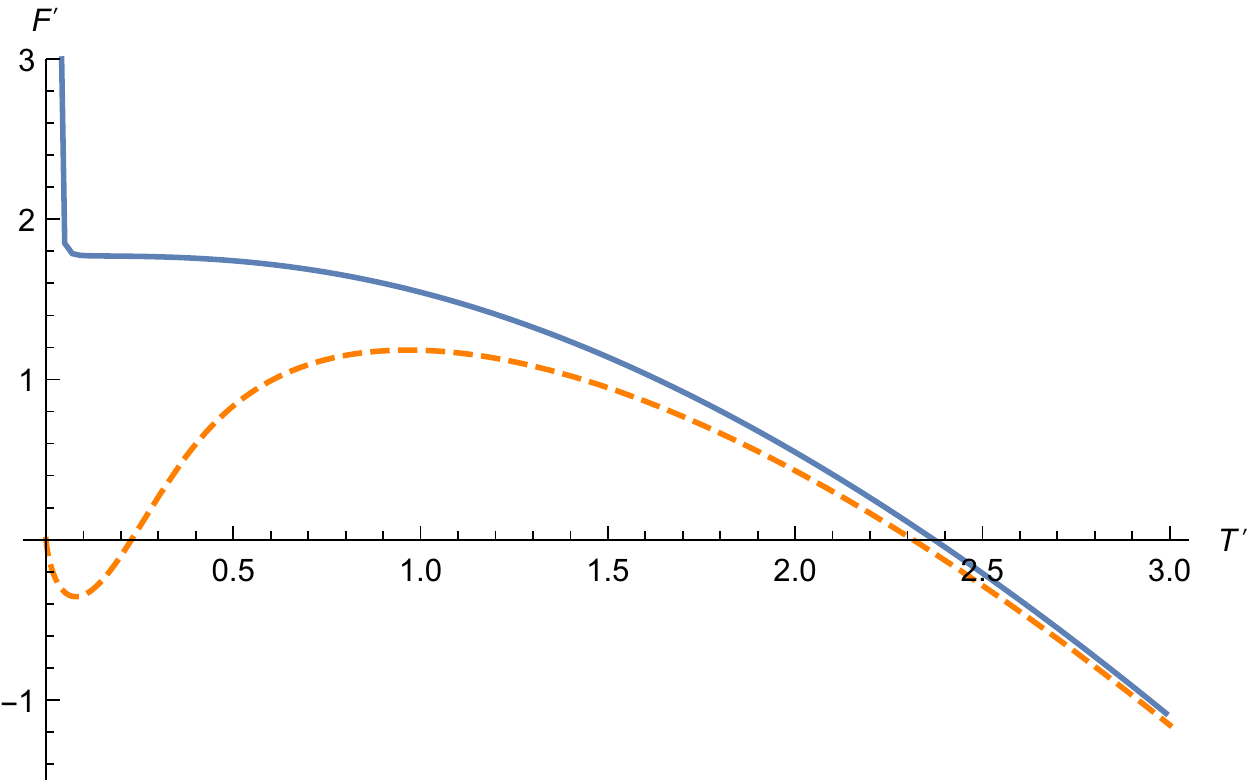}}
\hskip5mm
\subcaptionbox{Free energy up to $g=3$ \label{sfig:F3}}{\includegraphics[width=0.3\linewidth]{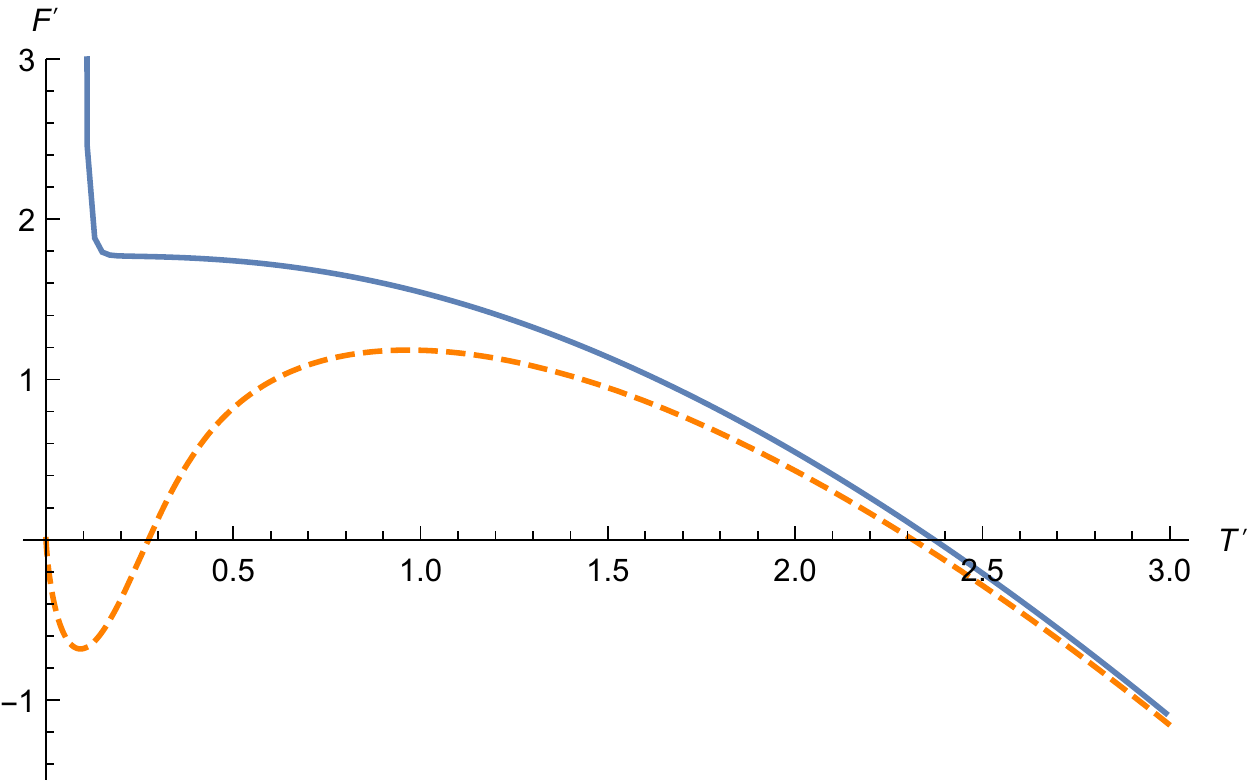}}
  \caption{
Plot 
of free energy including the higher genus corrections up to \subref{sfig:F1} $g=1$, \subref{sfig:F2} $g=2$ and
\subref{sfig:F3} $g=3$.
The blue solid curves are the quenched free energy in
\eqref{eq:que-Z[g]} while the orange dashed curves represent
the annealed free energy $-T'\log\bra Z\ket_{[g]}$.
}
  \label{fig:F123}
\end{figure}

In Fig.~\ref{sfig:F1}, \ref{sfig:F2} and
\ref{sfig:F3}, we show the plot of quenched free energy including the higher genus corrections
up to $g=1$, $g=2$ and $g=3$, respectively 
(see blue solid curves in Fig.~\ref{fig:F123}).
For $g=2$ and $g=3$, we see that the quenched free energy
diverges as $T\to0$.
This diverging behavior at low temperature comes from the fact that
$\cZ_2$ and $\cZ_3$ in \eqref{eq:Zg-result} become negative at some value of $W$.
This implies $\cZ_{[g]}$ in \eqref{eq:Z[g]} diverges in the negative direction
at some value of $W$ as $T\to0$.
As $\cZ_{[g]}\to-\infty$, the factor $e^{-\cZ_{[g]}}$ in \eqref{eq:que-Z[g]}
grows exponentially and it leads to the diverging behavior of the quenched free energy
in Fig.~\ref{sfig:F2} and
\ref{sfig:F3}.

However, this is not the expected behavior for the generating function
$\cZ(x)$.
In fact, since $Z$ and $x$ are positive quantities,
the generating function $\bra e^{-Zx}\ket$ satisfies the inequality
\begin{equation}
\begin{aligned}
 e^{-\cZ(x)}=\bra e^{-Zx}\ket<1,
\end{aligned} 
\label{eq:cZ-cond}
\end{equation}
which implies that $\cZ(x)$ is always positive.
The appearance of the negative coefficients in \eqref{eq:Zg-result}
simply means that the genus expansion is not a good approximation
of $\cZ(x)$ at low temperature since the expansion parameter $\xi$
is not small at low temperature $T\ll \hbar^{2/3}$.
In other words, the genus expansion of $\cZ(x)$ breaks down at low temperature.

To study the low temperature behavior of $\cZ(x)$ we need a different approximation
of $\cZ(x)$, which we will consider in section \ref{sec:low}.

\subsection{$\cZ_g$ from constitutive relation}\label{sec:constitutive}
Before discussing the low temperature behavior of $\cZ(x)$
in section \ref{sec:low},
let us consider an alternative method for the computation of
$\cZ_g$. This method clarifies the appearance of 
the Lambert function in this problem.
Using the boundary creation operator $\h{Z}$ in \eqref{eq:hatZ},
the generating function $\cZ(x)$ is obtained by acting
$e^{-\h{Z}x}$ to the topological free energy \eqref{eq:cZ-shift1}
\begin{equation}
\begin{aligned}
 \cZ(x)&=\cF(\{t_k\})-e^{-\h{Z}x}\cF(\{t_k\})=\cF(\{t_k\})-\cF(\{t_k'\}),
\end{aligned}
\label{eq:cZ-shift} 
\end{equation}
where $t_k'$ is given by
\begin{equation}
\begin{aligned}
t_k'&=t_k-x\hbar\rt{\frac{\bt}{\pi}}\bt^k.
\end{aligned} 
\label{eq:tk'}
\end{equation}
In the Airy case $t_k=0~(k\geq0)$, the first term of \eqref{eq:cZ-shift}
vanishes and hence we have
\begin{equation}
\begin{aligned}
 \cZ(x)=-\cF\Bigl(\Bigl\{-x\hbar\rt{\frac{\bt}{\pi}}\bt^k\Bigr\}\Bigr).
\end{aligned} 
\label{eq:cZ-shift2} 
\end{equation} 

To see the structure of the derivatives of $\cZ_g$ with respect to $t_0,t_1$,
it is convenient to work in the subspace \eqref{eq:off-shell}
with non-zero $t_0,t_1$.
On this subspace \eqref{eq:off-shell},
the genus-zero string equation \eqref{eq:string} for the shifted coupling $t_k'$
\eqref{eq:tk'} becomes
\begin{equation}
\begin{aligned}
 v=\sum_{k=0}^\infty t_k'\frac{v^k}{k!}=t_0+t_1v-x\hbar\rt{\frac{\bt}{\pi}}e^{\bt v}.
\end{aligned} 
\label{eq:string-shift}
\end{equation}
Here we denote $u_0$ for the background $t_k'$ \eqref{eq:tk'}
as $v$ in order to distinguish it from
$u_0$ in \eqref{eq:u01} for the original background \eqref{eq:off-shell}.
Now we see that
the string equation \eqref{eq:string-shift} for the background \eqref{eq:tk'}
can be written in the form of the Lambert function
\begin{equation}
\begin{aligned}
 z=We^{W},
\end{aligned} 
\label{eq:lambert}
\end{equation}
where $z$ is given by \eqref{eq:z-offshell} and $W$ is defined by
\begin{equation}
\begin{aligned}
W&=\frac{\bt t_0}{1-t_1}-\bt v.
\end{aligned}
\label{eq:W-v} 
\end{equation}
Namely, the Lambert function naturally appears from the genus-zero string equation for
the shifted background \eqref{eq:tk'}.

As shown in \cite{Eguchi:1994cx}, the genus-$g$ free energy
$\cF_g$ in \eqref{eq:cF-genus} is written as a combination of $v_m=\del_0^m v$,
which is known as the constitutive relation \cite{Dijkgraaf:1990nc}.
From \eqref{eq:lambert} and \eqref{eq:W-v}, one can show that
$v_m$ at $t_0=t_1=0$ is given by
\begin{equation}
\begin{aligned}
v_m=\cob_{m,1}-\bt^{m-1}D^mW,
\end{aligned} 
\label{eq:vm-DW}
\end{equation}
where $D$ is defined in \eqref{eq:def-D}.
In particular $v_1$ is given by
\begin{equation}
\begin{aligned}
 v_1=1-DW=\frac{1}{1+W}.
\end{aligned} 
\end{equation}
As discussed in \cite{Eguchi:1994cx}, if we assign the weight
$m-1$ to $v_m$, the genus-$g$ free energy $\cF_g$ has the weight $3g-3$.
From \eqref{eq:vm-DW}, this implies that $\cF_g$ is written as
\begin{equation}
\begin{aligned}
 \cF_g=\bt^{3g-3}\h{\cF}_g,
\end{aligned} 
\end{equation}
where $\h{\cF}_g$ is a function of $W$ only.
Then, in the Airy limit $t_0=t_1=0$
\eqref{eq:cZ-shift2} becomes
\begin{equation}
\begin{aligned}
 \cZ(x)=-\sum_{g=0}^\infty(\rt{2}\hbar)^{2g-2}\bt^{3g-3}\h{\cF}_g
=\sum_{g=0}^\infty \xi^{g-1}\cZ_g,
\end{aligned} 
\end{equation}
which implies
\begin{equation}
\begin{aligned}
 \cZ_g=-2^{g-1}\h{\cF}_g.
\end{aligned} 
\end{equation}
For instance, using the expression of
$\cF_g$ in terms of $v_m$ \cite{Eguchi:1994cx}
and plugging $v_m$ in \eqref{eq:vm-DW} into $\cF_g$,
we find
\begin{equation}
\begin{aligned}
 \cZ_1&=-\h{\cF}_1=-\frac{1}{24}\log v_1=\frac{1}{24}\log(1+W),\\
\cZ_2&=-2\h{\cF}_2=-2\bt^{-3}\left(\frac{v_2^3}{360v_1^4}
-\frac{7v_2v_3}{1920v_1^3}
+\frac{v_4}{1152v_1^2}\right)
=\frac{5W-19W^2+4W^3}{2880(1+W)^5}.
\end{aligned} 
\end{equation}
As expected, this agrees with the result \eqref{eq:Zg-result}
obtained from the recursion relation.
We have also checked that $\cZ_3$ in \eqref{eq:Zg-result}
is correctly reproduced from $\cF_3$ in \cite{Eguchi:1994cx}.

We note in passing that 
our $\cZ_g$ is related to the generating function of the Hodge integrals
studied in \cite{dubrovin2017classical}.\footnote{It is 
well-known that the Lambert function
also appears in the problem of Hurwitz numbers \cite{Bouchard:2007hi,Eynard:2009xe,mulase2009polynomial}.
As pointed out in \cite{dubrovin2017classical}, the generating function of 
Hurwitz numbers is a special case of the generating function of the Hodge integrals
(see the last page of \cite{dubrovin2017classical}
for details).}
For instance, $\cZ_2$ is expanded as
\begin{equation}
\begin{aligned}
 \cZ_2=-\frac{7}{720
   (1+W)^5}+\frac{11}{576
   (1+W)^4}-\frac{31}{2880
   (1+W)^3}+\frac{1}{720
   (1+W)^2}.
\end{aligned} 
\end{equation}
This agrees with $-2\cH_2^{\text{Hodge}}$ in 
eq.(87) of \cite{dubrovin2017classical}
after setting $x_i=0$ an $T=-W$.
From the result in \cite{dubrovin2017classical},
we find that $\cZ_g$ is expanded as
\begin{equation}
\begin{aligned}
 \cZ_g=\hf(1+W)^{2-2g}\sum_{n=1}^{3g-3}\ka_{g,n}\left(\frac{W}{1+W}\right)^n,
\end{aligned} 
\end{equation}
where
\begin{equation}
\begin{aligned}
 \ka_{g,1}=\frac{1}{12^gg!},\quad
\ka_{g,3g-3}=\frac{(-1)^gc_g}{12^g(5g-5)(5g-3)}.
\end{aligned} 
\end{equation}
As discussed in \cite{dubrovin2017classical}, 
$c_g$ is given by the coefficient of the formal power series
solution of the Painlev\'{e} I equation
\begin{equation}
\begin{aligned}
 U=\sum_{g=0}^\infty c_gX^{\hf(1-5g)},\quad
\frac{d^2U}{dX^2}+\frac{1}{16}U^2-\frac{1}{16}X=0.
\end{aligned} 
\end{equation}

\subsection{Low temperature resummation}\label{sec:low}
As shown in \cite{Okuyama:2019xvg},
at low temperature $\bra Z(\bt)^n\ket_c$ is approximated by
the one-point function $\bra Z(n\bt)\ket$ \eqref{eq:Znc-approx}
and hence the generating function $\cZ(x)$ is approximated by
\eqref{eq:cZ-lowsum}. The summation on the right hand side of 
\eqref{eq:cZ-lowsum} can be performed by rewriting
the one-point function $\bra Z(n\bt)\ket$ in terms of the eigenvalue
density $\rho(E)$
\begin{equation}
\begin{aligned}
 \bra Z(n\bt)\ket=\int_{-\infty}^\infty dE\rho(E)e^{-n\bt E}.
\end{aligned} 
\end{equation}
For the Airy case, $\rho(E)$ is given by
\begin{equation}
\begin{aligned}
 \rho(E)=\hbar^{-2/3}\left[\text{Ai}'(-\hbar^{-2/3}E)^2-
\text{Ai}(-\hbar^{-2/3}E)\text{Ai}''(-\hbar^{-2/3}E)\right],
\end{aligned} 
\end{equation}
where $\text{Ai}(z)$ denotes the Airy function.
We emphasize that this $\rho(E)$ is the exact eigenvalue density including all-genus
contributions. 
Then the low temperature approximation of $\cZ(x)$ in 
\eqref{eq:cZ-lowsum} becomes
\begin{equation}
\begin{aligned}
 \cZ(x)\approx -\sum_{n=1}^\infty \frac{(-x)^n}{n!}\int_{-\infty}^\infty dE\rho(E)e^{-n\bt E}
=\int_{-\infty}^\infty dE\rho(E)\Bigl(1-e^{-xe^{-\bt E}}\Bigr).
\end{aligned} 
\label{eq:cZ-low-app}
\end{equation}
This expression of $\cZ(x)$ is positive for $x>0$
since $\rho(E)$ is positive definite.
Thus this approximation of $\cZ(x)$ in \eqref{eq:cZ-low-app} 
satisfies the necessary condition \eqref{eq:cZ-cond}.

\begin{figure}[htb]
\centering
\includegraphics[width=0.45\linewidth]{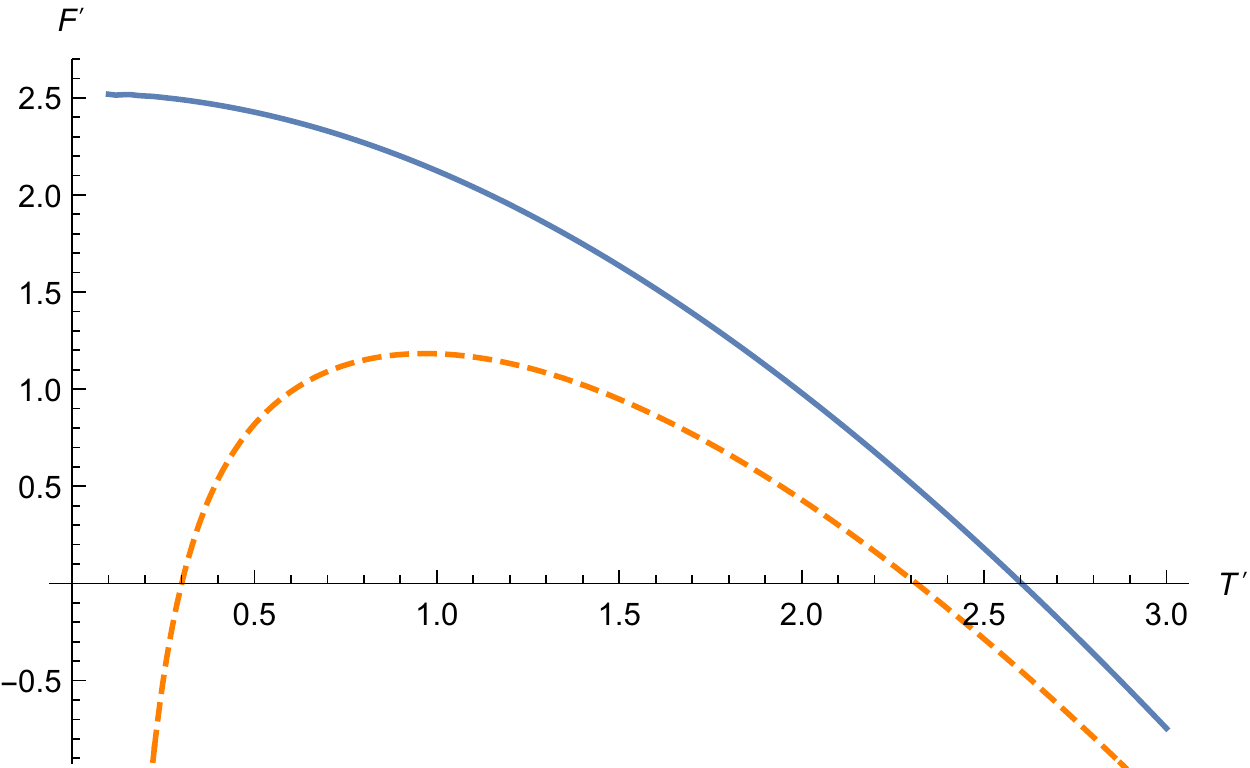}
  \caption{Plot of the free energy in the low temperature approximation.
The blue solid curve is the quenched free energy in the low temperature approximation
while the orange dashed curve is the all-genus result of the annealed free energy. 
}
  \label{fig:Frho-Airy}
\end{figure}

In Fig.~\ref{fig:Frho-Airy} we show the plot of
the quenched free energy using the low temperature approximation \eqref{eq:cZ-low-app}.
One can see that the quenched free energy is a monotonic function of
the temperature even in the low temperature regime.

\section{Conclusions and outlook}\label{sec:conclusion}
In this paper we have studied the quenched free energy
in an ensemble average of random systems.
We found the general formula \eqref{eq:formula} for the quenched free energy
and argued that it can be derived by the replica method 
\eqref{eq:replica} under a certain prescription
of the analytic continuation of $\bra Z^n\ket$.
We also discussed that the bulk gravity picture of our formula \eqref{eq:formula} 
involves the so-called spacetime D-branes $e^{-Zx}$
introduced in \cite{Marolf:2020xie}.
Then we have applied our general formula \eqref{eq:formula} to the Airy
limit of the Gaussian matrix model.
We found that the genus-zero approximation of the generating function
$\cZ(x)$ leads to a monotonic behavior of the quenched free energy.
The inclusion of higher genus corrections to $\cZ(x)$ leads to
a diverging free energy at low temperature, but this is not a problem 
of our formula \eqref{eq:formula};
it simply means that the genus expansion is not a good approximation 
at low temperature. Instead, we found that the alternative
low temperature approximation \eqref{eq:cZ-lowsum} gives rise to a 
well-behaved monotonic free energy even at low temperature.

There are many interesting open problems.
It would be interesting to generalize our analysis of the Airy case to
the matrix model of JT gravity \cite{Saad:2019lba}.
It is speculated in \cite{Saad:2019lba} that 
the quenched free energy of JT gravity exhibits 
a spin glass phase with the replica symmetry breaking at low temperature.
Based on the result in \cite{Okuyama:2019xbv,Okuyama:2020ncd},
it is argued in \cite{Engelhardt:2020qpv} that
in the low temperature scaling limit 
\begin{equation}
\begin{aligned}
 \bt\to\infty,\quad
S_0\to\infty\quad\text{with}\quad
\bt e^{2S_0/3}:~\text{fixed},
\end{aligned} 
\label{eq:Airy-lim}
\end{equation}
the quenched free energy of JT gravity reduces to that
of the Airy case.
Thus, according to the argument in \cite{Engelhardt:2020qpv},
we would be able to see the replica symmetry breaking
and spin glass phase in the quenched free energy of the Airy case
studied in our paper.
However, in our computation in section \ref{sec:Airy},
it is not obvious whether the replica symmetry breaking occurs in the Airy case or not.
Instead, our analysis is based on the spacetime D-branes
$e^{-Zx}$. It would be interesting to clarify
the relation between these two pictures.

Our analysis of the Airy case is limited to the genus expansion of $\cZ(x)$
in \eqref{eq:cZ-genus} and the low temperature approximation
of $\cZ(x)$ in \eqref{eq:cZ-lowsum}. 
It would be nice if we can find a closed form expression of the generating function
of the integral representation of $\bra Z^n\ket_c$ given 
in \cite{okounkov2002generating}.
We leave this as an interesting future problem.

In section \ref{sec:constitutive},
we have seen that $\cZ(x)$ in the Airy limit is closely related to the
generating function of the Hodge integrals studied in \cite{dubrovin2017classical}.
However, the quenched free energy is not the generating function of the Hodge integrals
itself, but some integral transformation of it.
By introducing the chemical potential $\mu=\log x$, \eqref{eq:formula2}
is written as
\begin{equation}
\begin{aligned}
 \bra\log Z\ket=\log\bra Z\ket-\int_{-\infty}^\infty
d\mu\Bigl[e^{-\cZ(e^\mu)}-e^{-\bra Z\ket e^\mu}\Bigr].
\end{aligned} 
\end{equation}
This reminds us of the frame change in the topological string/spectral theory 
(TS/ST) correspondence
(see e.g. \cite{Marino:2015nla} for a review).
It would be interesting to understand the mathematical meaning
of this integral transformation.

In our previous paper \cite{Okuyama:2020mhl}, 
we find that the quenched free energy of Gaussian matrix model
has a finite limit at zero temperature
\begin{equation}
\begin{aligned}
 \lim_{T\to0}F=E_0,
\end{aligned} 
\end{equation}
and $E_0$ is given by the average of the minimal eigenvalue of the random matrix. 
Our result in the Airy case suggests that the quenched free energy
in the Airy limit also has a finite limit at zero temperature 
(see Fig.~\ref{fig:Frho-Airy}).
It would be interesting to compute this zero temperature value of the free energy
and clarify its physical meaning.

Finally, let us comment on the status of our formula
\eqref{eq:formula2}.
We emphasize that \eqref{eq:formula2} is not just a formal expression, 
but it is useful in practice for the actual computation of
the quenched free energy in random matrix models.
We have explicitly demonstrated this computation for the Airy case
in the genus expansion and also 
in the low temperature approximation \eqref{eq:cZ-lowsum}.
For the JT gravity case, as shown in 
\cite{Okuyama:2019xbv,Okuyama:2020ncd}
the deviation from the Airy limit \eqref{eq:Airy-lim}
is organized as a low temperature expansion 
and it would be possible to compute $\cZ(x)$ of JT gravity in this expansion.
In general, the summation \eqref{eq:cZ-def} over the number of boundaries
might be an asymptotic series and one might worry that
$\cZ(x)$ is only defined perturbatively in the small $x$ expansion.
However, in principle $\cZ(x)$ can be defined non-perturbatively
as the free energy of random matrix model with 
a shifted background \eqref{eq:cZ-shift}, as demonstrated in section \ref{sec:constitutive} in the Airy case.
In other words, $\cZ(x)$ is the free energy of random matrix model
with a deformed matrix potential due to the insertion of the SD-brane operator 
$e^{-Zx}$. 
This definition of $\cZ(x)$ does not rely on the small $x$ expansion
in \eqref{eq:cZ-def} and it may serve as a non-perturbative definition of $\cZ(x)$, 
which in turn leads to a non-perturbative definition of $\bra\log Z\ket$
via \eqref{eq:formula2}.
Instead of computing the connected correlators $\bra Z^n\ket_c$ term by term in 
\eqref{eq:cZ-def}, we should consider the expectation value of the SD-brane operator
$\bra e^{-Zx}\ket$ as a whole.
We stress that 
this viewpoint is not obvious from the replica method \eqref{eq:replica}
and it is one of the advantages of our formalism \eqref{eq:formula2}, 
at least conceptually. 
One might think that the expansion \eqref{eq:young} used in the replica method
breaks down when $w_n$'s in \eqref{eq:wn} become of order one. This
is indeed the case at low temperature (see \eqref{eq:wn-order}).
However, the computation in section \ref{sec:replica} is just 
an illustration that our formula \eqref{eq:formula}
can be obtained from the replica method and the final
result \eqref{eq:formula} makes sense without referring to the replica
method. Our formula \eqref{eq:formula} does not require
$\bra \log Z\ket$ to be computed via the expansion \eqref{eq:young};
\eqref{eq:formula} is well-defined even at low temperature once $\bra e^{-Zx}\ket$
is defined non-perturbatively by a shift of background \eqref{eq:cZ-shift}. 
We hope that our expression of 
the quenched free energy \eqref{eq:formula} in terms of the SD-brane operator $e^{-Zx}$
provides us with a new perspective on our understanding of random systems 
and gravitational path integrals. 


\acknowledgments
The author would like to thank Kazuhiro Sakai for useful discussions.
This work was supported in part by JSPS KAKENHI Grant No. 19K03845.

\appendix
\section{High temperature expansion in the Airy limit}\label{app:high}
In this appendix we consider the 
high temperature expansion of the quenched free energy in the Airy limit.
The natural expansion parameter at high temperature is $\xi=\hbar^2\bt^3$.
For small $\xi$, $w_n$
defined in \eqref{eq:wn} behaves as
\begin{equation}
\begin{aligned}
 w_n=\cO(\xi^{n-1}).
\end{aligned} 
\label{eq:wn-order}
\end{equation}
Up to the order $\cO(\xi^2)$, the expansion \eqref{eq:young} reads
\begin{equation}
\begin{aligned}
 \frac{\bra Z^n\ket}{\bra Z\ket^n}=1+\hf n(n-1)w_2+\frac{1}{6}n(n-1)(n-2)w_3
+\frac{1}{8}n(n-1)(n-2)(n-3)w_2^2+\cO(\xi^3).
\end{aligned} 
\label{eq:Airy-wexp}
\end{equation}
For the Airy case, $w_2$ and $w_3$ are known in the closed form \cite{okounkov2002generating,Beccaria:2020ykg}. From \eqref{eq:airy-conn} we find
\begin{equation}
\begin{aligned}
 w_2&=\rt{\frac{\pi\xi}{2}}e^{\hf\xi}\text{Erf}\left(\rt{\frac{\xi}{2}}\right),\\
w_3&=\frac{4\pi\xi}{3\rt{3}}e^{2\xi}\Biggl[1-12T\Bigl(\rt{3\xi},\frac{1}{\rt{3}}\Bigr)\Biggr].
\end{aligned} 
\label{eq:wn-Airy}
\end{equation}
Plugging \eqref{eq:wn-Airy} into \eqref{eq:Airy-wexp},
we find the small-$\xi$ expansion of $\bra Z^n\ket/\bra Z\ket^n$
in the Airy limit
\begin{equation}
\begin{aligned}
 \frac{\bra Z^n\ket}{\bra Z\ket^n}=1+
\frac{n(n-1)}{2}\xi+\frac{n(n-1)(3n-5)(n+2)}{24}\xi^2+\cO(\xi^3).
\end{aligned} 
\end{equation}
This agrees with
the result eq.(1.17) in \cite{Beccaria:2020ykg}.
Taking the $n\to0$ limit \eqref{eq:replica}, we obtain the quenched free energy of the Airy case
as a small $\xi$ expansion
\begin{equation}
\begin{aligned}
 \bra \log Z\ket&=\log\bra Z\ket-\hf\xi+\frac{5}{12}\xi^2+\cO(\xi^3)\\
&=-\hf\log(4\pi\xi)-\frac{5}{12}\xi+\frac{5}{12}\xi^2+\cO(\xi^3),
\end{aligned} 
\end{equation}
where we used the result of one-point function $\bra Z\ket$ in \eqref{eq:airy-conn}.
For the higher $w_{n\geq4}$ we do not know the closed form expression, but the small
$\xi$ expansion of $w_n$ is easily obtained from the KdV equation
\eqref{eq:eq-cW}, as we will see below.

Let us briefly explain the computation of $w_n$ in the small $\xi$ expansion.
To do this, it is convenient to introduce the generating
function of the connected correlators
$\til{\cZ}(x)$, which is related to $\cZ(x)$ in \eqref{eq:cZ-def} by
\begin{equation}
\begin{aligned}
 \til{\cZ}(x)=-\cZ(-x)=\sum_{n=1}^\infty\frac{x^n}{n!}\bra Z^n\ket_c.
\end{aligned} 
\end{equation}
Then one can show that $\til{\cW}=\hbar\del_0\til{\cZ}$ satisfies a
similar relation as
\eqref{eq:eq-cW}
\begin{equation}
\begin{aligned}
 \del_1\til{\cW}=u\del_0\til{\cW}+\frac{\hbar^2}{6}\del_0^3\til{\cW}
+\hbar(\del_0\til{\cW})^2.
\end{aligned} 
\label{eq:tW-kdv}
\end{equation}
To solve this relation recursively, we work on the subspace \eqref{eq:off-shell}.
On this subspace \eqref{eq:off-shell}, the one-point function is given by
\begin{equation}
\begin{aligned}
 \bra Z(\bt)\ket=
\frac{1}{\rt{4\pi\hat{\xi}}}\exp\left(\frac{\hat{\xi}}{12}
+\frac{\bt t_0}{1-t_1}\right).
\end{aligned} 
\end{equation}
Here $\hat{\xi}=\hat{\hbar}^2\bt^3$ and $\hat{\hbar}$ is defined in
\eqref{eq:hat-hbar}.
From the result of $\bra Z^n\ket_c$ for small $n$ in \eqref{eq:airy-conn},
it is natural to make an ansatz
\begin{equation}
\begin{aligned}
 \bra Z(\bt)^n\ket_c=\bra Z(n\bt)\ket C_n(\hat{\xi}).
\end{aligned} 
\label{eq:ansatz}
\end{equation}
The important point is that $C_n$ is independent of $t_0$
and it depends on $t_1$ only through the combination  
$\hat{\hbar}$ in \eqref{eq:hat-hbar}.
From \eqref{eq:tW-kdv}, $W_n=\hbar\del_0\bra Z(\bt)^n\ket_c$ satisfies
\begin{equation}
\begin{aligned}
 \del_1W_n&=u\del_0W_n+\frac{\hbar^2}{6}\del_0^3W_n\\
&+\hbar\sum_{m=1}^{n-1}\binom{n}{m}\del_0W_m\del_0W_{n-m}.
\end{aligned} 
\label{eq:Wn-eq}
\end{equation}
Using the fact that the one-point part $\bra Z(n\bt)\ket$ of the ansatz \eqref{eq:ansatz}
satisfies the first line of \eqref{eq:Wn-eq}, we find the equation for
$C_n(\hat{\xi})$
\begin{equation}
\begin{aligned}
 (1-t_1)\del_1C_n=\rt{\frac{\hat{\xi}}{4\pi}}\sum_{m=1}^{n-1}
\binom{n}{m}\rt{nm(n-m)}e^{-\qu nm(n-m)\hat{\xi}}C_mC_{n-m}.
\end{aligned}
\label{eq:recC-t1} 
\end{equation}
Since $C_n$ depends on $t_1$ only through the combination $\hat{\xi}$,
the left hand side of \eqref{eq:recC-t1} is equal to $2\hat{\xi}\del_{\hat{\xi}}C_n$. 
Finally, after setting $t_0=t_1=0$, we find the recursion relation for $C_n(\xi)$
\begin{equation}
\begin{aligned}
 2\xi\del_\xi C_n=\rt{\frac{\xi}{4\pi}}\sum_{m=1}^{n-1}
\binom{n}{m}\rt{nm(n-m)}e^{-\qu nm(n-m)\xi}C_mC_{n-m}.
\end{aligned} 
\label{eq:rec-C}
\end{equation}
One can easily solve this equation recursively starting from $C_1=1$.
For instance, $C_2$ satisfies
\begin{equation}
\begin{aligned}
 \del_\xi C_2=\frac{e^{-\hf \xi}}{\rt{2\pi\xi}}.
\end{aligned} 
\end{equation}
This is solved as
\begin{equation}
\begin{aligned}
 C_2=\text{Erf}\left(\rt{\xi/2}\right),
\end{aligned} 
\end{equation}
which agrees with
the known result \eqref{eq:airy-conn} of the two-point function
in the Airy limit.
In a similar manner, for $n=3$ 
we can show that 
\begin{equation}
\begin{aligned}
 C_3=1-12T(\rt{3\xi},1/\rt{3})
\end{aligned} 
\end{equation}
is the solution for
the recursion relation \eqref{eq:rec-C}.

For $n\geq4$ we do not have a closed form solution for $C_n$, but it
is not difficult to solve the recursion relation
\eqref{eq:rec-C} in the small $\xi$ regime
and find $C_n$ as a power series in $\xi$.
Then $w_n=\bra Z^n\ket_c/\bra Z\ket^n$ is also obtained as
a power series in $\xi$
\begin{equation}
\begin{aligned}
 w_n=n^{-\frac{3}{2}}(4\pi\xi)^{\frac{n-1}{2}}e^{\frac{n^3-n}{12}\xi}C_n(\xi).
\end{aligned} 
\end{equation}
For instance, $w_4$ and $w_5$ are expanded as
\begin{equation}
\begin{aligned}
 w_4&=32 \xi^3+84 \xi^4+\frac{1936\xi^5}{15}+\cO(\xi^6),\\
w_5&=400 \xi^4+\frac{5776 \xi^5}{3} +\cO(\xi^6).
\end{aligned} 
\end{equation}
 
In this way, we finally find the small $\xi$ expansion of the quenched free energy
in the Airy limit of Gaussian matrix model
\begin{equation}
\begin{aligned}
 \bra \log Z\ket=-\hf\log(4\pi\xi)-\frac{5}{12}\xi+
\frac{5}{12}\xi^2-\frac{133}{90}\xi^3+\frac{8053}{840}\xi^4-\frac{171781}{1890}\xi^5+
\cO(\xi^6).
\end{aligned} 
\end{equation}
We have computed this expansion up to $\cO(\xi^{50})$.\footnote{The data of this expansion are available upon request to the author.}

\bibliography{paper}
\bibliographystyle{utphys}

\end{document}